\documentclass{article}
\usepackage{dcase2024,amsmath,graphicx,url,times,booktabs, tabularx,dsfont}
\usepackage{comment}
\usepackage{xcolor}
\usepackage{hyperref, url}
\usepackage{dsfont}
\usepackage{amssymb}
\usepackage{amsmath}
\usepackage{array}
\usepackage{booktabs}
\usepackage{siunitx}
\usepackage{pifont}
\usepackage{marvosym}

\title{Improving Query-By-Vocal Imitation \\
with Contrastive Learning and Audio Pretraining}

\name{Jonathan Greif$^{1}$, Florian Schmid$^1$, Paul Primus$^1$, Gerhard Widmer$^{1,2}$}
\address{$^1$Institute of Computational Perception (CP-JKU), $^2$LIT Artificial Intelligence Lab,\\ 
        Johannes Kepler University Linz, Austria \\
        \{jonathan.greif, florian.schmid, paul.primus\}@jku.at\\ 
 }

\begin{document}

\ninept
\maketitle

\begin{sloppy}

\begin{abstract}
Query-by-Vocal Imitation (QBV) is about searching audio files within databases using vocal imitations created by the user's voice.  Since most humans can effectively communicate sound concepts through voice, QBV offers the more intuitive and convenient approach compared to text-based search. 
To fully leverage QBV, developing robust audio feature representations for both the vocal imitation and the original sound is crucial. 
In this paper, we present a new system for QBV that utilizes the feature extraction capabilities of Convolutional Neural Networks pre-trained with large-scale general-purpose audio datasets. We integrate these pre-trained models into a dual encoder architecture and fine-tune them end-to-end using contrastive learning. A distinctive aspect of our proposed method is the fine-tuning strategy of pre-trained models using an adapted NT-Xent loss for contrastive learning, creating a shared embedding space for reference recordings and vocal imitations. The proposed system significantly enhances audio retrieval performance, establishing a new state of the art on both coarse- and fine-grained QBV tasks\footnote{Code at: \url{https://github.com/Jonathan-Greif/QBV}}.



\end{abstract}

\begin{keywords}
audio retrieval, vocal imitation, dual encoder, contrastive learning, QBV 
\end{keywords}

\vspace{-3pt}
\section{Introduction}
\label{sec:intro}
\vspace{-4pt}
Traditional audio retrieval systems rely on textual descriptions or keywords to search for audio recordings (e.g., \cite{koepke_journal,huang_audio_retrieval,primus_eusipco2024, primus_dcase2023}). Those descriptors are well suited to describe acoustic events on a high level. However, conveying specific acoustic nuances, such as pitch, loudness, timbre, or temporal relationships, via textual descriptions is difficult. For example, non-experts might struggle to find the right vocabulary to describe specific computer-synthesized sound effects.
However, since most humans can effectively imitate acoustic events through their vocal tract, Query-by-Vocal Imitation (QBV) becomes an attractive alternative.
In fact, previous work has suggested that QBV-based search engines actually achieve higher user satisfaction than text-based search engines~\cite{VROOM}. 

\begin{figure}[ht]
\vspace{-6pt}
    \centering
    \def\svgwidth{1 \linewidth}
\begingroup%
  \makeatletter%
  \providecommand\color[2][]{%
    \errmessage{(Inkscape) Color is used for the text in Inkscape, but the package 'color.sty' is not loaded}%
    \renewcommand\color[2][]{}%
  }%
  \providecommand\transparent[1]{%
    \errmessage{(Inkscape) Transparency is used (non-zero) for the text in Inkscape, but the package 'transparent.sty' is not loaded}%
    \renewcommand\transparent[1]{}%
  }%
  \providecommand\rotatebox[2]{#2}%
  \newcommand*\fsize{\dimexpr\f@size pt\relax}%
  \newcommand*\lineheight[1]{\fontsize{\fsize}{#1\fsize}\selectfont}%
  \ifx\svgwidth\undefined%
    \setlength{\unitlength}{283.46456693bp}%
    \ifx\svgscale\undefined%
      \relax%
    \else%
      \setlength{\unitlength}{\unitlength * \real{\svgscale}}%
    \fi%
  \else%
    \setlength{\unitlength}{\svgwidth}%
  \fi%
  \global\let\svgwidth\undefined%
  \global\let\svgscale\undefined%
  \makeatother%
  \begin{picture}(1,0.6)%
    \lineheight{1}%
    \setlength\tabcolsep{0pt}%
    \put(0,0){\includegraphics[width=\unitlength,page=1]{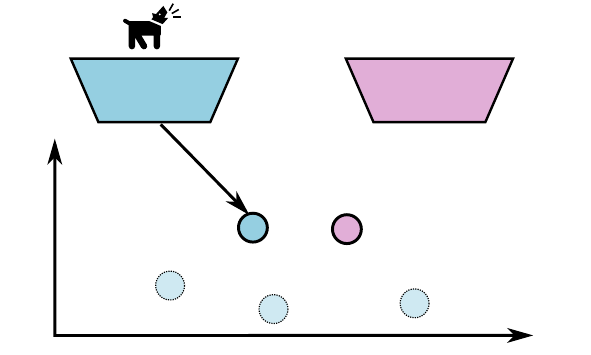}}%
    \put(0.74974541,0.51512321){\makebox(0,0)[lt]{\lineheight{1.25}\smash{\begin{tabular}[t]{l}woof\end{tabular}}}}%
    \put(0,0){\includegraphics[width=\unitlength,page=2]{siamese.pdf}}%
    \put(0.72666084,0.55575434){\makebox(0,0)[lt]{\lineheight{1.25}\smash{\begin{tabular}[t]{l}woof\end{tabular}}}}%
    \put(0.03457924,0.32878887){\makebox(0,0)[lt]{\lineheight{1.25}\smash{\begin{tabular}[t]{l}$d_1$\end{tabular}}}}%
    \put(0.83751328,0.06585815){\makebox(0,0)[lt]{\lineheight{1.25}\smash{\begin{tabular}[t]{l}$d_2$\end{tabular}}}}%
    \put(0.20044596,0.43783012){\makebox(0,0)[lt]{\lineheight{1.25}\smash{\begin{tabular}[t]{l}$\phi_a(a_i)$\end{tabular}}}}%
    \put(0.68030563,0.43783012){\makebox(0,0)[lt]{\lineheight{1.25}\smash{\begin{tabular}[t]{l}$\phi_v(v_i)$\end{tabular}}}}%
    \put(0,0){\includegraphics[width=\unitlength,page=3]{siamese.pdf}}%
  \end{picture}%
\endgroup%

    \caption{Two separate audio embedding models $\phi_a$ and $\phi_v$ project the reference sounds $a$ and the vocal imitations $v$ into a shared metric space. The contrastive loss increases the similarity of vocal imitations and their corresponding 
    sounds while pushing mismatching pairs away from each other in this metric space.}
    \label{fig:overview}
\vspace{-12pt}
\end{figure}
Previous work on QBV systems such as TL-IMINET~\cite{SiameseNetworkTL} and M-VGGish~\cite{MVGGish} relied on custom or relatively outdated pretrained audio embedding models and simple (non-contrastive) loss functions for training. Pishdadian et al. \cite{VimSketch} showed that simple signal processing methods based on handcrafted features outperformed those systems in their experimental setups~\cite{VimSketch}. 
In this work, we leverage contrastive training and the feature extraction capabilities of a more recent, pre-trained Convolutional Neural Network (CNN) model in a dual encoder architecture to improve QBV. The approach is sketched in Figure~\ref{fig:overview}. Experiments conducted on VimSketch~\cite{VimSketch} and VocalImitationSet~\cite{VocalImitationSet} demonstrate that our method outperforms the previous deep-learning-based approaches and the handcrafted approach
(Sections \ref{subsec: coarse} and \ref{subsec: fine}). We further conducted an ablation study in Section~\ref{subsec: ablation} to measure the impact of each of our proposed method's design choices.

\vspace{-4pt}
\section{Related Work}
\label{sec: format}
\vspace{-4pt}

Query-by-vocal Imitation (QBV) is a special case of \mbox{Query-by-Example}~(QBE)~\cite{QBE}. QBE encompasses various audio retrieval tasks such as cover song recognition~\cite{coversong}, query-by-beatboxing~\cite{beatbox}, and query-by-humming~\cite{humming1, humming2}. Unlike these music-related applications, QBV specifically focuses on general sound search. 

Among the most recent advancements in QBV are systems like TL-IMINET~\cite{SiameseNetworkTL} and CR-IMINET~\cite{VROOM}. Those are based on CNN-based dual encoder architectures, which rely on two separate embedding towers for the two domains (real and imitated sounds). Instead of comparing the embedding vectors directly, \cite{VROOM,SiameseNetworkTL} incorporate a Feedforward Neural Network~(FNN) that takes the embedding vectors as input and outputs an estimate of their similarity. 
TL-IMINET distinguishes itself by employing transfer learning, while CR-IMINET incorporates a Recurrent Neural Network layer. 
Another noteworthy system used for QBV is \mbox{M-VGGish}~\cite{MVGGish}, which combines features extracted from intermediate layers of VGGish~\cite{VGGish}. The model was pre-trained for audio tagging but not fine-tuned with 
imitations and reference sounds  
for QBV. \mbox{M-VGGish} assesses similarities by measuring the cosine similarity between the embedding vectors. On the \mbox{VocalSketch}~\cite{Vocalsketch} dataset
M-VGGish demonstrated superior performance compared to TL-IMINET, highlighting the feature extraction capabilities of models pre-trained on large audio tagging datasets~\cite{MVGGish}. 
However, recent work showed that these latest QBV systems perform poorly compared to simple signal processing~(SP) methods in certain settings~\cite{VimSketch}. The most performant of these SP methods involved converting the signals into the frequency domain using the Constant-Q Transform (CQT) and further with a 2D Fourier transformation (2DFT). The resulting representations were then compared using the cosine similarity. 



\vspace{-4pt}
\section{Proposed System}
\label{sec:pagelimit}
\vspace{-4pt}

Similarly to previous methods~\cite{SiameseNetworkTL,VROOM,MVGGish}, our system relies on two separate audio embedding models to project reference sounds $a_i$ and vocal imitations $v_i$ thereof, into a shared embedding space (see Figure~\ref{fig:overview}). In the following, we will denote the model that is used to embed the reference and the imitated acoustic events as $\phi_a$ and $\phi_v$, respectively. The correspondence between a vocal imitation $v_i$ and a reference sounds $a_i$ is determined by their distance in the embedding space.  Our proposed system improves over previous deep-learning-based QBV solutions in two main aspects, namely, the audio embedding model and the fine-tuning strategy on reference sound and imitation pairs using contrastive learning. In Section~\ref{subsec:audio_embedding_model} we motivate the choice for the audio embedding model and in Section~\ref{subsec:contrastive} we describe how these models are fine-tuned using contrastive learning to align vocal imitations and reference sounds in the shared embedding space.

\vspace{-4pt}
\subsection{Audio Embedding Model}
\label{subsec:audio_embedding_model}
\vspace{-4pt}

Extracting high-quality audio embeddings is a fundamental building block of a well-performing QBV system. The quality of these embeddings extracted by deep learning systems depends both, on the neural network architecture and the audio dataset it has been trained on. Previous QBV systems used small, custom architectures (e.g., TL-IMINET~\cite{SiameseNetworkTL} or CR-IMINET~\cite{VROOM}), or architectures that are outdated from today's point of view (M-VGGish~\cite{MVGGish}). 
These architectures were either directly trained on vocal imitation-reference sound pairs~\cite{VROOM}, pre-trained on smaller domain-specific datasets, and then fine-tuned on vocal imitation-reference sound pairs~\cite{SiameseNetworkTL}, or pre-trained on larger audio tagging datasets~\cite{MVGGish} but not fine-tuned on vocal imitation-reference sound pairs.

Our approach uses MobileNetV3 (MN)~\cite{MobileNetV3}, a modern efficient CNN pre-trained on AudioSet~\cite{Audioset}, as an audio embedding model. AudioSet is a large general-purpose audio dataset, consisting of 2 million 10-second audio clips labeled with 527 sound event classes. MNs achieve highly competitive performance on AudioSet when trained with Knowledge Distillation~\cite{knowledge_distilation} from a large transformer ensemble~\cite{efficientMN}. Additionally, pre-trained MNs have been shown to extract high-quality audio embeddings across music, environmental sound, and speech tasks~\cite{LowComplexAudioEmb}. We hypothesize that the general audio feature extraction capabilities obtained from AudioSet pre-training renders MN a strong choice for both the reference sound ($\phi_a$) and the vocal imitation ($\phi_v$) tower in our dual encoder setup. 

\vspace{-4pt}
\subsection{Contrastive Learning}
\label{subsec:contrastive}
\vspace{-4pt}
Typical training datasets consist of $N$ pairs, each holding a recording of a reference sound and a corresponding vocal imitation, i.e.,  $\{(a_i,~v_i)\}_{i=1}^N$.
During training, the two audio embedding networks learn to map inputs into a shared $D$-dimensionl space in which vocal imitations live close to their corresponding reference sounds. 
This alignment is achieved through contrastive training, which brings the embeddings of matching pairs ($a_i$,~$v_i$) together while pushing the representations of non-matching pairs ($a_i$,~$v_{j;j\neq i}$) apart. The correspondence between a vocal imitation $v_i$ and reference sound $a_i$ is determined using the cosine similarity between the embedded vectors in the shared embedding space:

\begin{equation}
S_{i j}=\frac{\phi_{\mathrm{a}}\left(a_i\right)^T \cdot \phi_{\mathrm{v}}\left(v_j\right)}{\left\|\phi_a\left(a_i\right)\right\|^2\left\|\phi_v\left(v_j\right)\right\|^2}.
\end{equation}
If each imitation corresponds to exactly one reference sound and vice versa, then the similarity matrix $S \in \mathbb{R}^{N\times N}$ contains the agreement scores for matching pairs along its diagonal, while the off-diagonal elements represent the agreement scores for mismatching pairs. 
A popular loss function for contrastive training that has not been explored in the QBV context yet is the NT-Xent~\cite{NT-Xent} loss. This loss first converts these similarities into a probability distribution over reference sounds via a temperature-scaled softmax activation. It then minimizes the cross entropy between the estimated distribution and a target distribution. In our case, the target distribution puts the entire probability mass on the reference recording for a given vocal imitation.
The corresponding loss is then defined as follows: 

\begin{equation}
\vspace{-5pt}
\mathcal{L}=-\frac{1}{N}\sum_{j=1}^N \log \frac{\exp \left(S_{j j} / \tau\right)}{\sum_{i=1}^N \exp \left(S_{i j} / \tau\right)\mathds{1}_{i\neq j}},
\vspace{3pt}
\end{equation}
where $\tau$ is a temperature hyper-parameter.

\vspace{-4pt}
\section{Experimental Setup}
\label{sec:pagestyle}
\vspace{-4pt}

We investigated our system's ability to retrieve the correct audio recording on two levels of granularity: coarse-grained and fine-grained. The corresponding experimental setups are explained below. This section further details the audio embedding model, the training procedure, and the evaluation metrics.

\vspace{-4pt}
\subsection{Coarse-grained QBV}
\label{subsec: coarse}
\vspace{-4pt}

In the course-grained setup, we evaluate the system's ability to recognize acoustic events (e.g., "dog barking," "paper tearing," or "thunderstorm") and correctly connect them across the two domains. The retrieved audios only need to contain the same event as the imitation in this setup to count as a match; specific acoustic properties like pitch, loudness, timing, etc. are not required to match.

We relied on the same experimental setup as in \cite{VROOM} to make our results comparable to theirs. To this end, we trained and evaluated our method on the VimSketch~\cite{VimSketch} dataset. This dataset contains 542 reference sounds and between 13 and 37 corresponding vocal imitations for each of them. As described in \cite{VROOM}, we only used 528 reference sounds and their corresponding imitations and split them into 10 folds, each containing around 52 sound events, for cross-validation.
Since the reference sounds mostly belong to distinct categories (i.e., two reference sounds typically don't share the same acoustic event), this setup is well-suited to measure the system's coarse-grained retrieval abilities.

\vspace{-4pt}
\subsection{Fine-grained QBV}
\label{subsec: fine}
\vspace{-4pt}

In the fine-grained setup, we evaluate the system's ability to retrieve a specific audio recording from a set of candidates that all contain the same acoustic event, e.g., its ability to select the best matching dog bark from a diverse collection of dog barks.

We relied on the same experimental setup as Kim et al.~\cite{VocalImitationSet} to compare our method to theirs, i.e., we trained the proposed system on \mbox{VocalSketch} v1.0.4~\cite{Vocalsketch} and evaluated it on VocalImitationSet~\cite{VocalImitationSet}.
\mbox{VocalSketch} v1.0.4 includes 240 unique reference sounds and around 18 corresponding vocal imitations for each of them; we used half of the data set for training and the other half for validation. Since the exact training-validation split used in \cite{VocalImitationSet} has not been made public, we randomly split the data according to their criteria. 
VocalImitationSet includes 302 unique reference sounds and around 18 vocal imitations for each. Those imitations were created to match their corresponding reference sounds exactly. In addition, each reference sound is also associated with approximately nine hard negative examples that contain the same acoustic event but differ with respect to other acoustic qualities. 
We relied on these hard negative examples to asses the systems' abilities to find exact matches among the multiple similar candidate recordings.

\vspace{-6pt}
\subsection{Embedding Networks}
\vspace{-4pt}

As discussed in Section~\ref{subsec:audio_embedding_model}, we chose efficient MobileNetV3~\cite{MobileNetV3}, pre-trained on AudioSet~\cite{Audioset}, as our embedding model. Specifically, we use a publicly available checkpoint referred to as \texttt{mn10\_as} (available via GitHub\footnote{\url{https://github.com/fschmid56/EfficientAT}}) because it strikes a good balance between computational efficiency and performance on the AudioSet benchmark. For audio pre-processing, we match the original feature extraction pipeline of the pre-trained MN~\cite{efficientMN} for both the reference sounds and the vocal imitations. Furthermore, we truncated or zero-padded all files to a duration of 10 seconds, aligned with MN's AudioSet pre-training setup. 

\vspace{-6pt}
\subsection{Training \& Augmentations}
\vspace{-4pt}

We used the Adam~\cite{Adam} as an optimizer featuring a learning rate schedule that includes an exponential warm-up (4 epochs), a constant phase (4 epochs), a linear decrease (14 epochs), followed by a fine-tuning phase (8 epochs). We trained for 30 epochs in total with a batch size of 16. 
The learning rate was set to 5e-4 and 7e-5 in the coarse-grained and fine-grained training setups, respectively. 
For the NT-Xent loss, we chose a temperature value of $\tau=0.07$.

To prevent overfitting, we applied multiple data augmentations on vocal imitations and reference sounds during training. We relied on the following methods:
\vspace{-4pt}
\begin{itemize}
    \item Time shifting: We randomly shift the waveform forward or backward within a range of 4000 steps.
    \vspace{-2pt}
    \item Time masking: The mel-spectrogram representations were randomly time-masked with a maximum length of 400 steps.
    \vspace{-2pt}
    \item Frequency masking: The mel-spectrogram representations were randomly frequency masked with a maximum of 4 bins.
    \vspace{-2pt}
    \item Freq-MixStyle~\cite{Mixstyle}: Frequency bands in spectrograms were normalized and denormalized again with mixed frequency statistics of other spectrograms from the same batch. With a probability of 0.3, Freq-MixStyle is applied to a batch and the mixing coefficient was drawn from a Beta distribution $B(0.4, 0.4)$.

\end{itemize}




\vspace{-4pt}
\subsection{Metric}
\vspace{-4pt}
Aligned with \cite{SiameseNetworkTL, VocalImitationSet, MVGGish, VimSketch, VROOM}, we assessed the retrieval performance with Mean Reciprocal Rank~(MRR)~\cite{MRR}:
\begin{equation}
\vspace{-4pt}
M R R=\frac{1}{|Q|} \sum_{i=1}^{|Q|} \frac{1}{\text { rank }_i}, 
\vspace{-0pt}
\end{equation}
where $\text{rank}_i$ denotes the rank of the target sound among all sounds for the i-th vocal imitation query; Q represents the number of imitations. MRR values range from 0 to 1, with higher values indicating better retrieval performance. In addition to the MRR and aligned with \cite{VocalImitationSet} we report the Mean Recall@k for k=1 and k=2 (MR@1 \& MR@2). This metric reflects the proportion of queries that successfully retrieved the target sound within the top k items in the search results. 

\vspace{-4pt}
\section{Results \& Discussion}
\label{sec:typestyle}
\vspace{-4pt}

This section presents the retrieval performance of our proposed system for coarse- and fine-grained QBV and a comparison to selected systems from the related work.
We additionally conducted an ablation study to understand the impact of our design choices better.

\begin{table*}[h!]
\vspace{-15pt}
	\centering
	\begin{tabular}{l|c|cc|ll|ccc}
     \toprule
    \multicolumn{1}{c}{ } & \multicolumn{1}{c}{ } & \multicolumn{2}{c}{\textbf{Supervised Pre-Training}} & \multicolumn{2}{c}{\textbf{Contrastive Fine-Tuning}} & \multicolumn{3}{c}{\textbf{Performance}} {} \\
    \textbf{Model} & Dual & AudioSet & VimSketch   & Loss & Similarity & MRR & MR@1 &MR@2\\
    \midrule
    MN & \checkmark &   \checkmark       & - &  - & cos & 0.295 & 0.175 & 0.258\\
    MN & \checkmark &    \checkmark       & -   & BCE & FNN &  0.354 & 0.183 & 0.306\\
    MN & \checkmark &    \checkmark       & -   & BCE & cos &  0.439 & 0.26 & 0.43\\
    \midrule
    MN & \checkmark &     \checkmark         & \checkmark &  - & cos & 0.508 & 0.35& 0.497\\
    MN & \checkmark &    \checkmark       & \checkmark &   BCE & cos  & 0.582 & 0.422 & 0.595 \\
    \midrule 
    MN & \checkmark &    \checkmark       & \checkmark   & NT-Xent & cos & 0.614 & 0.463 & 0.619\\
    MN & \checkmark &    \checkmark       & - &  NT-Xent & cos & \textbf{0.635} & \textbf{0.478} & \textbf{0.649}\\
    MN & \checkmark &   -       & - &  NT-Xent & cos & 0.493 & 0.322 & 0.477\\
    \midrule
    MN & - & \checkmark    &\checkmark & NT-Xent& cos & 0.553 & 0.399 & 0.544 \\
    MN & - &\checkmark & - &  NT-Xent & cos & 0.575 & 0.411 & 0.583\\
    \bottomrule
	\end{tabular}
 	\caption{Ablation study of design choices on the coarse-grained QBV setting. \textit{Dual} refers to using shared or separate encoders for reference sounds and imitations; \textit{Supervised Pre-training} indicates whether the encoders were pre-trained on the class labels in \textit{AudioSet} and/or \textit{VimSketch} (vocal imitations only); \textit{Loss} indicates which loss was used for contrastive fine-tuning ('-' in this column means no training on reference--imitation pairs); \textit{Similarity} of two embeddings was either measured with cosine similarity (\textit{cos}) or with an \textit{FNN}. }
  \label{tab:abl}
  \vspace{-6pt}
\end{table*}

\vspace{-4pt}
\subsection{Coarse-grained QBV}
\vspace{-4pt}

We compared results on the coarse-grained benchmark for \mbox{M-VGGish}, 2DFT, and MN to the results for \mbox{CR-IMINET} and \mbox{TL-IMINET} reported in \cite{VROOM}. Table \ref{tab:results1} shows the results. 
Our method achieved the highest MRR of 0.631, substantially outperforming both hand-crafted approaches like 2DFT~\cite{VimSketch} (0.308) and previous deep-learning-based methods like \mbox{CR-IMINET}~\cite{VROOM} (0.348), \mbox{TL-IMINET}~\cite{SiameseNetworkTL} (0.325) and \mbox{M-VGGish}~\cite{MVGGish} (0.228).

We note that TL-IMINET performed much better than M-VGGish and 2DFT, contrary to previously reported results~\cite{MVGGish, VimSketch}. This is likely due to the larger training set (476 vs.~120 reference sounds), which benefits models that are trained on imitation and reference pairs (i.e., TL-IMINET) but not those that are not (i.e., M-VGGish, 2DFT).

\begin{table}[h!]
	\centering
	\begin{tabular}{c|l|l|l}
		\toprule
		\textbf{Model} & \textbf{MRR} & \textbf{MR@1}  & \textbf{MR@2}\\
        \midrule    
            CR-IMINET* & 0.348 $\pm$ 0.03 & - & - \\
            TL-IMINET*& 0.325 $\pm$ 0.03 &  - & -\\
            M-VGGish & 0.228 $\pm$ 0.016 & 0.118 $\pm$ 0.018 & 0.182 $\pm$ 0.018\\ 
            2DFT of CQT& 0.309 $\pm$ 0.021 & 0.169 $\pm$ 0.016 & 0.268 $\pm$ 0.025 \\ 
            MN (ours) & \textbf{0.631} $\pm$ 0.027& \textbf{0.479} $\pm$ 0.031  & \textbf{0.646} $\pm$ 0.034 \\ 
		\bottomrule
	\end{tabular}
 	\caption{Results for coarse-grained evaluation on VimSketch as described in Section~\ref{subsec: coarse}; ranges give the standard deviation across the ten folds. (*Results taken from \cite{VocalImitationSet})}
  \label{tab:results1}
\vspace{-4pt}
\end{table}

\vspace{-4pt}
\subsection{Fine-grained QBV}
\vspace{-4pt}

\begin{table}[h!]
	\centering
	\begin{tabular}{c|c|c|c}
		\toprule
		\textbf{Model} & \textbf{MRR} & \textbf{MR@1}  & \textbf{MR@2}\\
        \midrule
            TL-IMINET* & 0.356 & 0.151 & 0.278\\
            M-VGGish & 0.416 & 0.212 & 0.364\\ 
            2DFT of CQT & 0.489 & 0.293 & 0.451\\ 
            MN (ours) & 0.476& 0.278 & 0.449\\ 
            MN (ours)\textsuperscript{\Cross} & \textbf{0.513}& \textbf{0.313} & \textbf{0.493}\\ 
		\bottomrule
	\end{tabular}
 	\caption{Results for the fine-grained evaluation on VocalImitationSet as described in Section~\ref{subsec: fine}. * denotes results taken from \cite{VocalImitationSet} and \textsuperscript{\Cross} indicates that supervised pre-training on vocal imitations is used.} 
  \label{tab:results2}
  \vspace{-8pt}
\end{table}

The performance on the fine-grained benchmark is shown in Table~\ref{tab:results2}.
We also experimented with training the AudioSet pre-trained MN further with vocal imitations in the training dataset by predicting their corresponding sound classes, as an additional training phase before the contrastive learning stage. When doing so, our proposed system outperformed previous methods and achieved the highest MRR of 0.513, surpassing TL-IMINET (0.356), M-VGGish (0.416), and 2DFT (0.489). Nevertheless, the margin of the signal processing method is smaller compared to coarse-grained QBV. When omitting the supervised pre-training, the performance of MN (0.476) falls behind that of 2DFT. 
This indicates that the granularity of the embedding space should be further improved to allow better discrimination between recordings that contain the same acoustic event. Since our method was not explicitly trained to distinguish between fine-grained details in recordings belonging to the same concept, we hypothesize that optimizing for such a scenario will result in further performance gains.

\vspace{-4pt}
\subsection{Ablation Study}
\label{subsec: ablation}
\vspace{-4pt}

Our ablation study, detailed in Table \ref{tab:abl}, demonstrates the effectiveness of our proposed method's components in the coarse-grained setting. 

By comparing the MN embeddings of reference sounds and imitations with cosine similarity (without additional training on reference--imitation datasets), our system achieved 0.295 MRR (row 1), which is similar to 2DFT (see Table~\ref{tab:results1}). 

This setting also allows a comparison between the pre-trained audio embedding models; when the MN is replaced with VGGish, the MRR dropped from 0.295 to 0.228 (compare row 1 in Table~\ref{tab:abl} and row 3 in Table~\ref{tab:results1}). This confirms our hypothesis that MN is the stronger choice for the dual encoder setup.

Interestingly, fine-tuning the AudioSet pre-trained MN on vocal imitations (as described in Section~\ref{subsec: fine}) resulted in an MRR increase of more than 0.2 (from 0.295 to 0.508) without any contrastive training involved (compare the first rows in section 1 \& 2 of Table~\ref{tab:abl}).

TL- and CR-IMINET used an FNN with a single output and a Binary Cross Entropy~(BCE) loss to learn the similarity between two embeddings. We tried the same with our architecture, which only resulted in a relatively small improvement (from 0.295 in row 1 to 0.354 MRR in row 2 in Table~\ref{tab:abl}). Replacing the FNN with cosine similarity improved the MRR further to 0.439 (row 3 of Table~\ref{tab:abl}), indicating that using the FNN head is not beneficial.


Replacing the BCE loss with the NT-Xent loss increased the performance substantially, e.g., from 0.439 to 0.635 MRR when no supervised training on vocal imitations was used (compare row 3 in section 1 and row 2 in section 3) and from  0.582 to 0.614 with supervised training on vocal imitations (compare row 2 in section 2 and row 1 in section 3). This is likely because the NT-Xent loss relies on multiple negative examples in each update.

Interestingly, supervised pre-training with vocal imitations decreased performance in combination with pre-training on AudioSet, NT-Xent loss, and cosine similarity (compare rows 1 \& 2 in section 3). This indicates that AudioSet pre-training is more beneficial for coarse-grained retrieval, whereas model parameters additionally pre-trained on vocal imitations in a supervised fashion allow a more fine-grained distinction (as demonstrated by the results for fine-grained QBV in Table~\ref{tab:results2}). 


Using a shared embedding network for vocal imitations and reference recordings led to a performance drop (see Section 4). This is consistent with results reported by Zhang et al.\;\cite{SiameseNetworkTL}, who also suggest that specialized encoders for the two domains are better suited for feature extraction. 


Overall, the results indicate that the combination of pre-training on AudioSet with two independent embedding networks and contrastive training with NT-Xent loss 
enhances the retrieval accuracy of QBV systems.

\vspace{-4pt}
\section{Conclusion}
\label{sec:page}
\vspace{-4pt}

This paper proposes a Query-by-Vocal Imitation system that improves upon previous approaches by integrating a modern, efficient CNN, pre-trained on large-scale AudioSet, in a dual encoder setup. The encoders are fine-tuned using contrastive learning with an adapted NT-Xent loss, aligning vocal imitations with their reference recordings in a shared embedding space. Our results demonstrate that the proposed system substantially enhances retrieval performance, establishing a new state of the art on both coarse- and fine-grained QBV tasks. Unlike previous deep learning-based solutions, the presented system clearly outperforms manually extracted features. We believe that our proposed system represents a step forward towards integrating QBV into sound search engines, ultimately making it easier and more intuitive to search for sounds.

\vspace{-4pt}
\section{ACKNOWLEDGMENT}
\label{sec:ack}
\vspace{-4pt}

The LIT AI Lab is supported by the Federal State of Upper Austria. Gerhard Widmer's work is supported by the European Research Council (ERC) under the European Union's Horizon 2020 research and innovation programme, grant agreement No 101019375 (Whither Music?).
\bibliographystyle{IEEEtran}
\bibliography{refs}

%
%
%
%
%
%
%
%
%

\end{sloppy}

\end{document}